\documentclass[12pt,a4paper]{article}
\usepackage{amsmath}
\usepackage{amsxtra}
    \usepackage{amstext}
    \usepackage{amssymb}
    \usepackage{latexsym}
    \usepackage{graphicx}
\usepackage{color}
\usepackage{graphics}
\usepackage{cite}

\newcommand{\ce}{{\cal E}}
\newcommand{\cp}{{\cal P}}

\newcommand{\p}{\vspace{6pt}\noindent}
\newcommand{\jump}{\vspace{2pt}}

\advance\textheight by \topskip
\makeatletter

\@addtoreset{equation}{section}
\def\section{\@startsection {section}{1}{\z@}{-8.5ex plus -1ex minus
 -.2ex}{3.3ex plus .2ex}{\large\bf}}%\centering}}
\def\subsection{\@startsection{subsection}{2}{\z@}{-3.25ex plus
 -1ex minus -.2ex}{1.5ex plus .2ex}{\bf}}
\def\subsubsection{\@startsection{subsubsection}{3}{\z@}{-3.25ex plus%
 -1ex minus -.2ex}{1.5ex plus .2ex}{\sl}}

\begin{document}

\begin{titlepage}
\vspace*{-2cm}
\begin{flushright}
%%YITP-05-29
\end{flushright}

\vspace{0.3cm}

\begin{center}
{\Large {\bf Type II defects revisited}} \\
\vspace{1cm} {\large  E.\ Corrigan\footnote{\noindent E-mail: {\tt edward.corrigan@york.ac.uk}}}\\
\vspace{0.5cm}
{\em Department of Mathematics \\ University of York, York YO10 5DD, U.K.} \\
\vspace{0.3cm} {\large and}\\ \vspace{0.5cm}
{\large C.\ Zambon\footnote{\noindent E-mail: {\tt cristina.zambon@durham.ac.uk}}} \\
\vspace{0.3cm}
{\em Department of Physics \\ Durham University, Durham DH1 3LE, U.K.} \\

\vspace{2cm} {\bf{ABSTRACT}}\\ \end{center}   \vspace{.5cm}

\p Energy and momentum conservation in the context of  a type II, purely transmitting, defect, within a single scalar relativistic two-dimensional field theory, places a severe constraint not only on the nature of the defect but also on the potentials for the scalar fields to either side of it. The constraint is of an unfamiliar type since it requires the Poisson Bracket of the defect contributions to energy and momentum with respect to the defect discontinuity and its conjugate to be balanced by the potential difference across the defect. It is shown that the only solutions to the constraint correspond to the known integrable field theories.

\p \\

\vfill
\end{titlepage}

\section{Introduction}

Defects within (relativistic) integrable field theory models in two dimensions have been studied for some time from both classical and quantum viewpoints (see, for example \cite{dms1994,kl1999,Mintchev02,bcz2003,bcz2004, bcz2005,gyz2006,cz2007,hk2007,gyz2007,c2008, cz2009,n2009,ad2012,agsz2014,d2016}). In essence, a defect always involves a discontinuity of some kind, and in an integrable model experience has shown that this discontinuity is a jump in the field value at a specific point (similar to the discontinuity in velocity across a shock in a fluid flow), with `sewing' conditions across the defect relating the fields on either side in such a manner that suitably adjusted conservation laws are maintained. Characteristically, such defects break space translation invariance and are purely transmitting. Intriguingly, insisting upon sewing conditions that maintain the conservation of energy and momentum seems to be sufficient to guarantee integrability. There is no direct proof of this but there is a body of evidence from many specific cases that indicates it should be the case.

\p So far, there are basically two types of defect that appear to be integrable, called type I (where the defect has no degrees of freedom of its own \cite{bcz2003,bcz2004}), and type II (where the defect carries its own degrees of freedom \cite{cz2009,cr2013}). However, they can be mixed together as they have been recently, for example, to discuss defects within the $d^{(1)}_r$ series of affine Toda field theories \cite{br2017}. There may be other possibilities, yet to be found, that for example encompass affine Toda field theories based on the $e_r^{(1)},\ r=6,7,8$ root systems.

\p The aim of this paper is to take the first step at a systematic classification, by examining defect sewing relations required to preserve energy-momentum conservation, but without specifying the field theories themselves, in order to determine the constraints on the field theory potentials. For type I this is straightforward and was carried out previously demonstrating that, for example, within the class of affine Toda field theories only those based on the roots represented by the extended Dynkin diagrams for $a^{(1)}_r$ can support type I defects \cite{bcz2004,cz2007}. The simplest example of the type I defect  is included here for comparison with the type II defect, which is more intricate. For type II, the analysis seems to be  far from straightforward and the main part of the paper is classifying the possibilities in the simplest of cases where there is a single scalar field defined on each side of the defect. In either situation, the only possiblities are the known integrable field theories except that the Tzitz\'eica model ($a_2^{(2)}$ affine Toda) is excluded from the set of models supporting type I defects but can support a type II defect.

\p One intriguing possibility is that integrable models are actually characterised by their ability to support integrable discontinuities. However, a proof of that fact, if true, remains distant.

\section{The formalism}

  In this paper, field theories will be analysed by examining carefully the sewing conditions across a defect taking into account the requirements of energy-momentum conservation including contributions from the defect. For the purposes of this article the defect is taken to be situated at $x=0$ (though in principle it might be situated anywhere along the $x$-axis), with scalar fields $u$ and $v$ to the left and right of it, respectively. There is no a priori assumption that the fields are of the same type, though they often are. In other words, in their respective domains the two fields satisfy the equations
$$\partial^2 u=-U^\prime(u),\ \  (x<0);\quad \partial^2 v=-V^\prime(v),\ \ (x>0),$$ where $U(u)$ and $V(v)$ are the potentials.
The field equations need to be supplemented by conditions relating the fields $u,v$  and/or their derivatives across the defect. The idea is that by making very few assumptions, not only the defect conditions are specified by the requirements, but also the potentials $U,\ V$.

\subsection{Energy}

\p Consider first the contributions to the total energy and how it might be conserved. The time derivative of the contributions to the total energy from the fields to either side of the defect is given (on using the equations of motion) by
$$\dot E=\int_{-\infty}^0\,\left(\frac{1}{2}(u_t^2+u_x^2)+U(u)\right)_tdx +\int_0^{\infty}\,\left(\frac{1}{2}(v_t^2+v_x^2)+V(v)\right)_tdx = [u_tu_x]^0+[v_tv_x]_0,$$
assuming the contributions from $\pm \infty$ are zero. Thus, the  sewing conditions should be designed to convert the right hand side to a total time derivative of the energy contributed by the defect.

\p One possibility (type I) is to require:
$$x=0:\quad u_x=v_t- {\cal E}_u,\ \ v_x=u_t+{\cal E}_v,$$
where ${\cal E}$ depends on both $u$ and $v$ and partial derivatives with respect to $u$ or $v$ are denoted by subscripts, then
$$\dot E=-u_t{\cal E}_u-v_t{\cal E}_v=-\frac{d{\cal E}}{dt}.$$
Thus, the total energy $E+{\cal E}$ is conserved.

\p Another possiblity (type II) is to introduce a quantity $\lambda$, defined only at $x=0$ but depending on time, and then to set
$$x=0:\quad u_x=\lambda_t- {\cal E}_u,\ \ v_x=\lambda_t+{\cal E}_v,\ \ u_t-v_t =-\ce_\lambda,$$
where now $\ce$ depends on $u,\ v$ and $\lambda$, then
$$\dot E=-u_t{\cal E}_u-v_t{\cal E}_v-\lambda_t\ce_\lambda=-\frac{d{\cal E}}{dt},$$
and $E+\ce$ is conserved as before, though in this case $\ce$ has additional dependence on $\lambda$. The defect does not break time translation invariance so it is not surprising that little effort is required to conserve energy, and the energy ${\cal E}$ introduced by the impurity is unconstrained.

\p It is also worth recalling that both sets of sewing relations follow directly from Lagrangian descriptions of the defect:
\begin{equation}\label{defectL}
{\cal L}(u,v)={\cal L}(u)\theta(-x)+{\cal L}_D\delta(x)+{\cal L}(v)\theta(x)
\end{equation}
with
\begin{equation}\label{defectLI}
{\cal L}(u)=\frac{1}{2}(u_t^2-u_x^2)-U(u),\quad {\cal L}(v)=\frac{1}{2}( v_t^2-v_x^2) -V(v),
\end{equation}
and with the type I or type II defect Lagrangian ${\cal L}_D$ given by
\begin{equation}\label{defectLII}
{\cal L}_I=uv_t-{\cal E}(u,v),\quad {\cal L}_{II}=(u-v)\lambda_t-{\cal E}(u,v,\lambda).
\end{equation}
In these expressions, subscripts denote derivatives with respect to $t$ and $x$,  and the defect energy functional ${\cal E}$ depends only on the fields not their time (or space) derivatives.
\subsection{Momentum}

\p In a similar manner, the time derivative of the contributions to the total field momentum is given by
$$\dot P=\int_{-\infty}^0\left(u_t u_x\right)_t dx +\int_0^{\infty}\left(v_t v_x\right)_t dx=\left[\frac{1}{2}(u_t^2+u_x^2)-U(u)\right]^0+\left[\frac{1}{2}(v_t^2+v_x^2)-V(v)\right]_0,$$
with the same assumption as before. Since space translation is broken explicitly by the defect the requirement of overall momentum conservation is expected to impose stringent conditions on the fields. The two cases introduced above will be dealt with separately.
\subsubsection{Type I}

\p Using the type I sewing conditions (in this section all fields are evaluated at $x=0$):
$$\dot P=-v_t\ce_u-u_t\ce_v +\frac{1}{2}\ce_u^2-\frac{1}{2}\ce_v^2 -U(u)+V(v)= -\frac{d\cp}{dt},$$
where $\cp$ is related to $\ce$ and strongly constrained by the following relationships:
\begin{equation}\label{typeIconditions}\ce_u=\cp_v,\ \ \ce_v=\cp_u,\ \ \frac{1}{2}\left(\ce_u^2-\ce_v^2\right) =U(u)-V(v).\end{equation}
These conditions are powerful. The first pair require that $\ce\pm\cp$ is a function of $u\mp v$. To examine the third condition, it is convenient to define new variables $p,\ q$ by
$$p=\frac{u+v}{2},\ \ q=\frac{u-v}{2},\ {\rm at}\  x=0,$$
then the last condition of \eqref{typeIconditions} becomes
$$\frac{\ce_p\,\ce_q}{2}=U(p+q)-V(p-q).$$
Then, since $\ce=F(p)+G(q), \ \cp=F(p)-G(q)$, for some functions $F,G$, this requires $$\frac{F^\prime(p)G^\prime(q)}{2}=U(p+q)-V(p-q),$$
which restricts possible choices for the potentials $U,V$. This is because the difference on the right hand side must factor into a function of $p$ multiplied by a function of $q$. From this observation, it is straightforward to find the possible solutions for $F,\ G, \ U,$ and $V$. It is enough to note that the left hand side must satisfy
$$\left(F^\prime(p)G^\prime(q)\right)_{pp}=\left(F^\prime(p)G^\prime(q)\right)_{qq}$$
and hence that
$$\frac{F^{\prime\prime\prime}}{F^\prime}=\frac{G^{\prime\prime\prime}}{G^\prime}=k^2,$$
where $k$ is constant. For example, if $k\ne 0$
$$F^\prime(p)=\alpha e^{kp}+\beta e^{-kp},\ \ G^\prime(q)=\gamma e^{kq}+\delta e^{-kq},$$
where $\alpha,\ \beta,\ \gamma,\ \delta$ are also constants. Also, if $k=0$ then
$$F^\prime(p)=\alpha p+\beta,\ \ G^\prime(q)=\gamma q+\delta.$$
 Hence, the allowed potentials can be deduced leading to the following possibilities: the fields $u,v$ can both be free massive (with the same mass), or free massless, or both be Liouville, or both be sine/sinh-Gordon (with the same parameters). Or, one of $u$ or $v$ could be free massless and the other could be Liouville. In the latter case both field theories are conformal.

\subsubsection{Type II}

\p Using the type II sewing conditions leads to a  different type of constraint on the defect contributions to the total energy and momentum. Considering the field contributions to the momentum, following the same steps as in the previous subsection, gives
$$\dot P=-p_t\ce_\lambda -\lambda_t\ce_p +\frac{1}{2}\ce_u^2-\frac{1}{2}\ce_v^2 -U(u)+V(v)=-\frac{d\cp}{dt},$$
which, assuming $\cp$ is a function only of $q,p,\lambda$, and noting
$$\frac{d\cp}{dt}=q_t\cp_q+p_t\cp_p+\lambda_t\cp_\lambda= -\left(\frac{1}{2}\ce_\lambda\cp_q-p_t\cp_p-\lambda_t\cp_\lambda\right),$$ requires
$$\ce_\lambda=\cp_p,\ \ \ce_p=\cp_\lambda, \ \ \frac{1}{2}\left(\cp_\lambda\ce_q-\cp_q\ce_\lambda\right)=U(p+q)-V(p-q).$$
The last of these is intriguing because as far as the defect contribution to the Lagrangian \eqref{defectLII} is concerned $\lambda$ and $q$ are conjugate variables. Thus, the nonlinear relationship states that the Poisson bracket with respect to these conjugate variables of the defect energy and momentum is twice the `potential difference' across the defect.

\p Now, since $\ce,\ \cp$ are functions of $\lambda,\ p, \ q$ and $\ce\pm\cp$ is a function of $p\mp \lambda$ together with $ q$, it follows that $$\ce=F(p+\lambda,q)+G(p-\lambda, q), \ \cp=F(p+\lambda,q)-G(p-\lambda,q).$$
 Then, explicitly in terms of $F,\ G$ the nonlinear Poisson Bracket constraint is:
\begin{equation}\label{PBrel}
F_\lambda G_q-F_q G_\lambda=\{F,G\}=U(p+q)-V(p-q).
\end{equation}
The constraint equation \eqref{PBrel} is powerful because the left hand side depends on $\lambda$ while the right hand side does not.

\p While several examples are known the general solution to \eqref{PBrel} is not yet clear. The objective in this article is to describe an approach to solving a functional equation of this unfamiliar type in which all four functions $F,G,U,V$ are strongly constrained. In particular, it is necessary to investigate whether or not there are any solutions beyond those known already all of which correspond to integrable field theories, namely, sine-Gordon, Tzitzi\'eca, Liouville and massive or massless free.

\section{An approach to solving the Poisson Bracket Equation}

\p One approach, used with success previously \cite{cz2009}, is to guess that solutions for $F,G$ must be sums of exponentials. An alternative might be to try to be systematic, assume each has a Taylor expansion and write
$$F(p+\lambda,q)=\sum_{k=0}^\infty \frac{(p+\lambda)^k}{2^k k!}\,f_k(q),\ \ G(p-\lambda,q)=\sum_{l=0}^\infty \frac{(p-\lambda)^l}{2^ll!}\,g_l(q).$$
Then the Poisson bracket relation becomes:
$$F_\lambda G_q-F_q G_\lambda=\frac{1}{2}\sum_{k,l} \frac{(p+\lambda)^k}{2^kk!}\frac{(p-\lambda)^l}{2^ll!}(f_{k+1}g^\prime_l+f_k^\prime g_{l+1}).$$
The latter can be rewritten (grouping together terms of constant $N=k+l$) as
$$F_\lambda G_q-F_q G_\lambda=\frac{1}{2}\sum_{N=0}^\infty\sum_{k=0}^N \frac{(p+\lambda)^k}{2^kk!}\frac{(p-\lambda)^{N-k}}{2^{N-k}(N-k)!}\,(f_{k+1}g^\prime_{N-k}+f_k^\prime g_{N-k+1}),$$
which seems to require the coefficients of each term in the set of terms corresponding to a particular $N$ to be the same (apart from the factorial factors)
so that gathering the terms together they can be recognised as being the coefficients in the binomial expansion of $(p+\lambda + p-\lambda)^N=(2p)^N$, which is clearly independent of $\lambda$, as required. In other words,
$$F_\lambda G_q-F_q G_\lambda=\frac{1}{2}\sum_{N=0}^\infty \frac{p^N}{N!}\, h_N(q),$$ where
\begin{equation}\label{fghrels}
h_N(q)=f^{\phantom{\prime}}_{k+1}\,g^\prime_{N-k}+f_k^\prime \, g^{\phantom{\prime}}_{N-k+1}, \ \ k=0,\dots, N.
\end{equation}

\p On the other hand, assuming the potentials also have a Taylor expansion, the right hand side can be written
\begin{equation}\label{}U(p+q)-V(p-q)=\frac{1}{2}\sum_{N=0}^\infty \frac{p^N}{N!}\,\left(U^{(N)}(q)-V^{(N)}(-q)\right)\end{equation}
where the superscript $(N)$ denotes the $N^{th}$ derivatives of $U,\ V$. Hence, formally,
$$h_N(q)=U^{(N)}(q)-V^{(N)}(-q).$$

\p The aim is to find compatible expressions for the $g,$ $f$ and $h$ sequences of  functions of $q$ in such a way as to be able to reconstruct the functions $G$ and $F$ and then seek potentials $U,$ $V$ such that \eqref{PBrel} will be satisfied.

\p In order to achieve this, assuming that none of the coefficients $f_k, g_k$ vanish,  notice first that the expressions \eqref{fghrels} can be rearranged to
\begin{equation}\label{fghrelsRearranged}
\frac{h_N}{f_{k+1}g_{N-k+1}}= a_{N-k+1}+b_{k+1},\ \  k=0,\dots, N, \quad N=0, 1,2,\dots\end{equation}
where
 \begin{equation}\label{abdefs}a_{N-k+1}=\frac{g'_{N-k}}{g_{N-k+1}},\ \ b_{k+1}=\frac{f'_{k}}{f_{k+1}}.
\end{equation}

\section{Special case: the sinh-Gordon model}
\label{h2k=0}
\p The simplest special case to analyse assumes the two potentials $U,V$ are the same and even. In other words $U(q)=V(q)=U(-q)$ and hence,   $h_N(q)=0$ when $N$ is even.
Then it follows directly from \eqref{fghrelsRearranged} and  \eqref{abdefs} that $a_l=-b_l$ for all positive integers $l.$ Moreover, $a_l=a_2$ if $l$ is an even positive integer and $a_l=a_1$ if $l$ is odd. Hence relations \eqref{fghrelsRearranged} can be rewritten to involve only the functions $a_1,$ $a_2$. Thus, for instance
\begin{eqnarray}\label{specialcase}
a_2-a_1&=&\frac{h_1}{f_1g_2}=-\frac{h_1}{f_2g_1},\quad N=1,\nonumber\\
a_2-a_1&=&\frac{h_3}{f_1g_4}=-\frac{h_3}{f_2g_3}=\frac{h_3}{f_3g_2}=-\frac{h_3}{f_4g_1},\quad N=3,
\end{eqnarray}
and so on.
Comparing these for different $N$ then implies relations among the  $f_l,g_l$ themselves. For example,
$$h_3=\frac{h_1 g_3}{g_1}=\frac{h_1 g_4}{g_2},\quad \Rightarrow \quad \frac{g_3}{g_1}=\frac{g_4}{g_2}.$$
On the other hand $a_1=a_3$ and $a_2=a_4,$ which implies
$$\frac{g_3}{g_1}=\frac{g_2'}{g_0'},\quad\frac{g_4}{g_2}=\frac{g_3'}{g_1'}.$$
It follows that
$$\frac{g_3}{g_1}=\frac{g_3'}{g_1'},$$
which in turn implies
$$g_3=\alpha\, g_1,$$
where $\alpha$ is a constant.
Hence,
$$g_2=\alpha g_0+c,\quad g_3=\alpha\, g_1,\quad g_4=\alpha^2\,g_0+\alpha c,\quad h_3=\alpha\, h_1,$$
where $c$ is also a constant.
Thus, in general:
$$g_{2k+1}=\alpha\, g_{2k-1},\quad g_{2k+2}=\alpha \, g_{2k},\quad h_{2k+1}=\alpha\, h_{2k-1},\quad k=1,\dots$$
implying
$$g_{2k+1}=\alpha^k\,g_1,\quad g_{2k}=\alpha^k\,g_0+\alpha^{k-1}c,\quad h_{2k+1}=\alpha^k\,h_1,\quad k=1,\dots\ .$$
In order to find expressions for members of the sequence of $\{ f_l\}$, remember that $b_1=-a_1$ and $b_2=-a_2,$ which implies
$$f_1=-\frac{f_0^\prime g_1}{g_0'},\quad f_2=-\frac{f_1^\prime g_2}{g_1'}.$$
On the other hand, the first line of \eqref{specialcase} requires
$$f_2=-\frac{f_1g_2}{g_1}.$$
Equating the two expressions for $f_2$ gives
$$\frac{f_1}{f_1'}=\frac{g_1}{g_1'}\quad \implies f_1=\beta \,g_1,$$
where $\beta$ is a constant. It follows that
$$f_0'=-\beta g_0',\quad f_2=-\beta(\alpha\,g_0+c),\quad f_3=\beta \alpha g_1,\quad f_4=-\beta \alpha (\alpha\,g_0+c),$$
and hence
$$f_{2k+1}=\beta\alpha^k\,g_1,\quad f_{2k}=-\beta(\alpha^k\,g_0+\alpha^{k-1}c).$$
Finally, still using \eqref{specialcase},
$$h_1=f_1g_2(a_2-a_1)=\beta\left(g_1g_1'-\alpha g_0g_0'-cg_0\right).$$
In summary:
\begin{eqnarray*}
g_{2k+1}&=&\alpha^k\,g_1,\quad  f_{2k+1}=\beta g_{2k+1},\quad k=0,1\dots\\
f_0&=&-\beta g_0+d,\quad g_{2k}=\alpha^k\,g_0+\alpha^{k-1}c,\quad f_{2k}=-\beta g_{2k},\quad k=1,2\dots,\\
h_1&=&\frac{\beta}{2}\left(g_1^2-\alpha g_0^2-2cg_0\right)', \quad h_{2k+1}=\alpha^k\,h_1,\quad k=0,1\dots.
\end{eqnarray*}
Here, $g_0,$ $g_1$ are undetermined functions of $q$, and $\alpha,$ $\beta,$ $c,$ and $d$ are constants.
Using this data, the reconstructed $F$ and $G$ functions are:
\begin{eqnarray*}
F(p+\lambda,q)&=&-\beta g_0\cosh\left(\frac{\sqrt{\alpha}(p+\lambda)}{2}\right)-\frac{\beta c}{\alpha}\left(\cosh\left(\frac{\sqrt{\alpha}(p+\lambda)}{2}\right)-1\right)\\
&&+\frac{\beta g_1}{\sqrt{\alpha}}\sinh\left(\frac{\sqrt{\alpha}(p+\lambda)}{2}\right)+d\\
G(p-\lambda,q)&=&g_0\cosh\left(\frac{\sqrt{\alpha}(p-\lambda)}{2}\right)+\frac{c}{\alpha}\left(\cosh\left(\frac{\sqrt{\alpha}(p-\lambda)}{2}\right)-1\right)\\
&&+\frac{g_1}{\sqrt{\alpha}}\sinh\left(\frac{\sqrt{\alpha}(p-\lambda)}{2}\right).
\end{eqnarray*}
Then
$$F_\lambda G_q-F_q G_\lambda=\frac{\beta}{2\sqrt{\alpha}}\left(g_1g_1'-\alpha g_0g_0'-cg_0'\right)\sinh(\sqrt{\alpha}\,p)=U(p+q)-U(p-q).$$
To satisfy this relation requires
$$g_1g_1'-\alpha g_0g_0'-cg_0'=A\,\left(e^{\sqrt{\alpha}\,q}-\,e^{-\sqrt{\alpha}\,q}\right),$$
from which it follows that
$$U(u)=\frac{A\beta}{4\sqrt{\alpha}}\,\left(e^{\sqrt{\alpha}\,u}+e^{-\sqrt{\alpha}\,u}\right),
$$
which correspond to the sinh-Gordon potential.

\p Note: if $\alpha=0$, then
$$F(p+\lambda,q)=-\frac{\beta c}{2}\left(\frac{p+\lambda}{2}\right)^2-\beta g_0+ d+\beta g_1 \left(\frac{p+\lambda}{2}\right),\quad
G(p-\lambda,q)=\frac{c}{2}\left(\frac{p-\lambda}{2}\right)^2 + g_0+ g_1 \left(\frac{p-\lambda}{2}\right),$$
and
$$F_\lambda G_q-F_q G_\lambda =U(p+q)-U(p-q)=\frac{\beta p}{2} \,(g_1g_1'-c g_0').$$
The only non zero solution to this requires
$$g_1g_1'-c g_0'\sim q,$$
and leads to the potential for a free massive scalar field,
$$U(u)=\frac{\beta}{8}\,u^2.$$

\section{General case}
\label{generalcase}

\p Consider a pair of relations \eqref{fghrelsRearranged}, which share one of the ratios appearing on their right hand side and then subtract them. If this operation is performed for all possible pairs sharing a common ratio,
the following expressions are found
 \begin{eqnarray}
f_{k+1}&=&\frac{h_{k+r}g_{s+1}-h_{k+s}g_{r+1}}{g_{r}'g_{s+1}-g_{s}'g_{r+1}},\nonumber\quad f_{k}'=\frac{h_{k+s}g_{r}'-h_{k+r}g_{s}'}{g_{r}'g_{s+1}-g_{s}'g_{r+1}},\quad r\neq s=0,1,\dots,\quad  k =0,1,\dots \nonumber\\
g_{k+1}&=&\frac{h_{k+r}f_{s+1}-h_{k+s}f_{r+1}}{f_{r}'f_{s+1}-f_{s}'f_{r+1}},\nonumber\quad g_{k}'=\frac{h_{k+s}f_{r}'-h_{k+r}f_{s}'}{f_{r}'f_{s+1}-f_{s}'f_{r+1}},\quad r\neq s=0,1,\dots,\quad  k =0,1,\dots \\\label{generalrelationsfg}
\end{eqnarray}
where formulas in the first and second lines are obtained taking into account `$f$-common' ratios and `$g$-common' ratios, respectively.

\p Consider first only a subset of these expressions. The strategy is to find a solution for the subset and look for a pattern that allows a generalisation of the formulas found
to all $f,$ $g,$ $h$-functions.
Then verify whether these expressions satisfy all \eqref{generalrelationsfg} relations or whether the remaining \eqref{generalrelationsfg} introduce further constraints on the possible solution.
The subset adopted contains the expressions, for which the pairs of indices $(r,s)$ in \eqref{generalrelationsfg} are $(0,1),$ $(1,2),$ $(0,2).$ Combine the two expressions obtained for each pair of indices $(r,s)$
by eliminating the $f$-functions. Then the relations found are:
 \begin{eqnarray}\label{GeneralRelations_hg_firstline}
g_{k+1}&=&\frac{\left|\begin{array}{ccc}
            h_k & h_{k+1} & 0 \\
            h_0 & h_1 & g_1 \\
            h_1 & h_2 & g_2
          \end{array}\right|}{H_A},\quad
g_{k}'=\frac{\left|\begin{array}{ccc}
            h_k & h_{k+1} & 0 \\
            h_0 & h_1 & g_0' \\
            h_1 & h_2 & g_1'
          \end{array}\right|}{H_A},\quad
g_{k+1}=\frac{\left|\begin{array}{ccc}
            h_{k+1} & h_{k+2} & 0 \\
            h_2 & h_3 & g_2 \\
            h_3 & h_4 & g_3
          \end{array}\right|}{H_B},\\ \nonumber \\
g_{k}'&=&\frac{\left|\begin{array}{ccc}
            h_{k+1} & h_{k+2} & 0 \\
            h_2 & h_3 & g_1' \\
            h_3 & h_4 & g_2'
          \end{array}\right|}{H_B},\quad
g_{k+1}=\frac{\left|\begin{array}{ccc}
            h_k & h_{k+2} & 0 \\
            h_0 & h_2 & g_1 \\
            h_2 & h_3 & g_3
          \end{array}\right|}{H_C},\quad
g_{k}'=\frac{\left|\begin{array}{ccc}
            h_k & h_{k+2} & 0 \\
            h_0 & h_2 & g_0' \\
            h_1 & h_3 & g_2'
          \end{array}\right|}{H_C},\label{GeneralRelations_hg_secondline}
\end{eqnarray}
where
\begin{eqnarray}\label{FGfunctions}
H_A&=&\left|\begin{array}{cc}
               h_1 & h_0 \\
               h_2 & h_1
             \end{array}
\right|=\left|\begin{array}{cc}
               g_0' & g_1 \\
               g_1' & g_2
             \end{array}
\right|\left|\begin{array}{cc}
               f_0' & f_1 \\
               f_1' & f_2
             \end{array}
\right|=G_AF_A,\nonumber\\
H_B&=&\left|\begin{array}{cc}
               h_3 & h_2 \\
               h_4 & h_3
             \end{array}
\right|=\left|\begin{array}{cc}
               g_1' & g_2 \\
               g_2' & g_3
             \end{array}
\right|\left|\begin{array}{cc}
               f_1' & f_2 \\
               f_2' & f_3
             \end{array}
\right|=G_BF_B,\nonumber\\
H_C&=&\left|\begin{array}{cc}
               h_2 & h_0 \\
               h_4 & h_2
             \end{array}
\right|=\left|\begin{array}{cc}
               g_0' & g_1 \\
               g_2' & g_3
             \end{array}
\right|\left|\begin{array}{cc}
               f_0' & f_1 \\
               f_2' & f_3
             \end{array}
\right|=G_CF_C.
\end{eqnarray}
The assumption is that $H_A$, $H_B$, $H_C$ are different from zero. The cases in which these determinants are zero do not lead to new results. An example of these cases will be discussed in appendix \ref{H_A=0}.

\p Look, for instance, at the expressions with $H_A$. It can be noticed that for $k=0$ and $k=1$ these relations are identically satisfied.
Additional information starts to emerge for $k=2.$ Similar
considerations can be applied to all the other expressions. Then, by expanding the determinants with respect to their $g$-column, the first non trivial relations from each expression in
\eqref{GeneralRelations_hg_firstline}, \eqref{GeneralRelations_hg_secondline} are:
\begin{equation}\label{noidentities-g}
g_3H_A=g_2\Lambda-g_1\Delta,\quad
g_1H_B=g_2\Gamma-g_3\Delta,\quad
g_2H_C=g_3\Lambda+g_1\Gamma,
\end{equation}
and
\begin{equation}\label{noidentities-gprime}
g_2'H_A=g_1'\Lambda-g_0'\Delta,\quad
g_0'H_B=g_1'\Gamma-g_2'\Delta,\quad
g_1'H_C=g_2'\Lambda+g_0'\Gamma,
\end{equation}
where
\begin{equation*}
\Lambda=\left|\begin{array}{cc}
           h_2 & h_3 \\
           h_0 & h_1
         \end{array}\right|,\quad
\Delta=\left|\begin{array}{cc}
           h_2 & h_3 \\
           h_1 & h_2
         \end{array}\right|,
         \quad
\Gamma=\left|\begin{array}{cc}
           h_2 & h_1 \\
           h_4 & h_3
         \end{array}\right|.
\end{equation*}
After some algebra, they lead to
\begin{equation}\label{g_2g_3_relations}
g_3=g_2\frac{\Lambda}{H_A}-g_1\frac{\Delta}{H_A},\quad g_2'=g_1'\frac{\Lambda}{H_A}-g_0'\frac{\Delta}{H_A},
\end{equation}
$$g_2\left(\Gamma H_A-\Lambda\Delta\right)=g_1\left(H_B H_A-\Delta^2\right),\quad
g_2\left(H_CH_A-\Lambda^2\right)=g_1\left(\Gamma H_A-\Lambda\Delta\right),$$
$$g_1'\left(\Gamma H_A-\Lambda\Delta\right)=g_0'\left(H_B H_A-\Delta^2\right),\quad
g_1'\left(H_C H_A-\Lambda^2\right)=g_0'\left(\Gamma H_A-\Lambda\Delta\right),$$
where the compatibility condition reads
$$\left(H_BH_A-\Delta^2\right)\left(H_CH_A-\Lambda^2\right)=\left(\Gamma H_A-\Lambda\Delta\right)^2.$$
Notice that it is possible to write the determinants $\Lambda,$ $\Delta$ and $\Gamma$ as products of $F$ and $G$-determinants \eqref{FGfunctions}. Better still,
the latter and the $H$-determinants as well can be written as products of only $F_A,$ $G_A,$ $G_B$ and $G_C.$ In fact
\begin{equation}\label{FGrelations}
H_A=\frac{F_A}{G_A}\,G_A^2,\quad H_B=\frac{F_A}{G_A}\,G_B^2,\quad H_C=\frac{F_A}{G_A}\,G_C^2,\quad
\frac{F_B}{F_A}=\frac{G_B}{G_A},\quad \frac{F_C}{F_A}=\frac{G_C}{G_A},
\end{equation}
$$\Lambda=\frac{F_A}{G_A}\,G_AG_C,\quad \Delta=\frac{F_A}{G_A}\,G_BG_A,\quad \Gamma=\frac{F_A}{G_A}\,G_CG_B.$$
It follows that
$$H_BH_A-\Delta^2=H_CH_A-\Lambda^2=\Gamma H_A-\Lambda\Delta=0$$
and also
$$H_BH_C-\Gamma^2=\Lambda H_B-\Delta\Gamma=\Delta H_C-\Lambda\Gamma=0,$$
which is equivalent to
\begin{eqnarray}\label{hDeterminant}
\left|\begin{array}{ccc}
  h_0 & h_1 & h_2 \\
  h_1 & h_2 & h_3 \\
  h_2 & h_3 & h_4
\end{array}\right|
=0.
\end{eqnarray}
The expressions \eqref{g_2g_3_relations} become
\begin{equation}\label{g3andg2prime}
g_3=g_2\frac{G_C}{G_A}-g_1\frac{G_B}{G_A},\quad g_2'=g_1'\frac{G_C}{G_A}-g_0'\frac{G_B}{G_A}.
\end{equation}
Finally from \eqref{hDeterminant} it possible to infer the following
$$
h_2=h_1\frac{G_C}{G_A}-h_0\frac{G_B}{G_A},\quad
h_3=h_2\frac{G_C}{G_A}-h_1\frac{G_B}{G_A},\quad
h_4=h_3\frac{G_C}{G_A}-h_2\frac{G_B}{G_A}.
$$
It seems that a pattern starts to emerge. In order to explore it, consider the expressions with $H_A$ in \eqref{GeneralRelations_hg_firstline} for $k=3.$ They lead to
\begin{equation}\label{g4andg3prime}
g_4=g_3\frac{G_C}{G_A}-g_2\frac{G_B}{G_A},\quad g_3'=g_2'\frac{G_C}{G_A}-g_1'\frac{G_B}{G_A}.
\end{equation}
On the other hand the last expression with $H_B$ in \eqref{GeneralRelations_hg_firstline} for $k=3$ and the middle expression with $H_A$ for $k=4$ lead to
\begin{equation}\label{h5andg4prime}
h_5=h_4\frac{G_C}{G_A}-h_3\frac{G_B}{G_A},\quad g_4'=g_3'\frac{G_C}{G_A}-g_2'\frac{G_B}{G_A}.
\end{equation}
Then,  differentiating expressions for $g_3,$ $g_4$ in \eqref{g3andg2prime}, \eqref{g4andg3prime} and comparing them with expressions for $g_3',$ $g_4'$ in \eqref{g4andg3prime}, \eqref{h5andg4prime}, it is found
$$g_2\left(\frac{G_C}{G_A}\right)'=g_1\left(\frac{G_B}{G_A}\right)',\quad g_3\left(\frac{G_C}{G_A}\right)'=g_2\left(\frac{G_B}{G_A}\right)'.$$
These expression are satisfied if $G_C/G_A$ and $G_B/G_A$ are constants or if
$$\frac{g_3}{g_2}=\frac{g_2}{g_1}\quad \Rightarrow \quad G_C=G_A\left(\frac{g_2}{g_1}\right)+ G_B\left(\frac{g_1}{g_2}\right)$$
and
$$\frac{g_1}{g_2}=\frac{G_C'G_A-G_A'G_C}{G_B'G_A-G_A'G_B}\quad \Rightarrow \quad G_A\left(\frac{g_2}{g_1}\right)'\left(G_A-G_B\left(\frac{g_2}{g_1}\right)^2\right)=0.$$
Since $G_A\neq 0$ and $(g_2/g_1)'\neq 0$ \footnote{In fact, $(g_2/g_1)'= 0$ implies $H_B=0,$ which also must be different from zero.}, it is found
$$\frac{G_B}{G_A}=\left(\frac{g_2}{g_1}\right)^2,\quad \frac{G_C}{G_A}=2\left(\frac{g_2}{g_1}\right),$$
where the latter is obtained using the expression for $g_3$ in \eqref{g3andg2prime}.

\p Before summarising the results obtained so far, a few words about the $f$-functions are necessary. Because of the symmetry between the $g$ and the $f$-functions, an analysis started
with expressions similar to the ones in \eqref{GeneralRelations_hg_firstline}, \eqref{GeneralRelations_hg_secondline} with the $g$-functions replaced by the $f$-functions, would have led to similar results. Note also the relations $F_C/F_A=G_C/G_A$ and $F_B/F_A=G_B/G_A$ in
\eqref{FGrelations}. Taking all of this into account, the tentative, uncompleted solutions are:
\begin{eqnarray}\label{solutionsI_II}
\mbox{Solution A:}\phantom{mmmmmm}\nonumber\\
\frac{G_C}{G_A}&=&\frac{F_C}{F_A}=2\,\xi,\quad \frac{G_B}{G_A}=\frac{F_B}{F_A}=\xi^2,\quad
\xi=\frac{g_2}{g_1},\quad f_2=f_1 \xi,\nonumber\\
g_0'&=&2\frac{g_1'}{\xi}-\frac{g_2'}{\xi^2},\quad f_0'=\frac{f_1'}{\xi}-\frac{f_1\xi'}{\xi^2}\nonumber\\
g_{k+1}&=&2\xi g_{k}-\xi^2g_{k-1},\ \ f_{k+1}=\xi^kf_1,\ \ h_k=2\xi h_{k-1}-\xi^2h_{k-2},\ k=2,3,\dots\nonumber\\
\end{eqnarray}
where $\xi$ is a function of $q$ and it cannot be a constant and
\begin{eqnarray}\label{solutionsI_I}
\mbox{Solution B:}\phantom{mmmmmm}\nonumber\\
\frac{G_C}{G_A}&=&\frac{F_C}{F_A}=a,\ \ \frac{G_B}{G_A}=\frac{F_B}{F_A}=b,
\nonumber\\
g_2&=&a \,g_1-b \,g_0+c,\ \
f_2=a \,f_1-b \,f_0+d,\nonumber\\
g_{k+1}&=&ag_{k}-bg_{k-1}, f_{k+1}=af_{k}-bf_{k-1}, h_k=ah_{k-1}-bh_{k-2},\ k=2,3,\dots\nonumber \\
\end{eqnarray}
where $a,$ $b,$ $c,$ $d$ are constants.

\subsection{Solution A}
\label{GeneralCaseII}

\p For solution A, the missing functions $h_0$ and $h_1$ can be found using \eqref{fghrels}. Then, the relations for solutions A can be rewritten in simple forms as
\begin{eqnarray}
g_0'&=&\left(\frac{g_1}{\xi}\right)',\quad f_0'=\left(\frac{f_1}{\xi}\right)',\quad
g_{k+1}=\xi^kg_1,\quad f_{k+1}=\xi^kf_1, \quad k=0,1,\dots\nonumber\\
\xi(q)&=&\frac{g_2}{g_1},\quad h_k=\xi^{k}\left(g_1\left(\frac{f_1}{\xi}\right)'+f_1\left(\frac{g_1}{\xi}\right)'+\frac{kf_1g_1\xi'}{\xi^2}\right),\quad k=0,1,\dots,
\end{eqnarray}
where $g_0',$ $g_1,$ $g_2,$ $f_0',$ $f_1$ are free functions of $q.$
These expressions can be used to verify that all \eqref{generalrelationsfg} relations are satisfied. Then, $G$ and $F$ functions can be reconstructed. They are:
$$F(p+\lambda,q)=f_0+\frac{f_1}{\xi}\left(e^{\xi(p+\lambda)/2}-1\right),\quad G(p-\lambda,q)=g_0+\frac{g_1}{\xi}\left(e^{\xi(p-\lambda)/2}-1\right).$$
It follows that
$$F_\lambda G_q-F_q G_\lambda=\frac{e^{p\,\xi }}{2}\left(f_1\left(\frac{g_1}{\xi}\right)'+g_1\left(\frac{f_1}{\xi}\right)'+p\frac{f_1g_1\xi'}{\xi}\right).$$
Given that the function $\xi$ cannot be a constant, there are no potentials $U$  and $V$ such that \eqref{PBrel} is satisfied.

\subsection{Solution B}
\label{GeneralCaseI}

Solution B seems to be more complicated to analyse. Similarly to what has been done for solution A, it is possible to obtain expressions for $h_0$ and $h_1$ using \eqref{fghrels}. They are:
\begin{eqnarray}
h_0&=&f_0'g_1+f_1g_0',\nonumber\\
h_1&=&g_1f_1'+g_0'(a f_1-b f_0 +d)=f_1g_1'+f_0'(a g_1-b g_0 +c).\label{h_0h_1I}
\end{eqnarray}
The second line provides a constraint. Before investigating this constraint, it is useful to look at the expressions for $h_2$ provided by \eqref{fghrels}, that is:
$$h_2=g_1f_2'+g_0'f_3=f_0'g_3+f_1g_2'=g_2f_1'+g_1'f_2.$$
Using \eqref{solutionsI_I}, it is easy to see that the first two expressions for $h_2$ can be rewritten as $h_2=ah_1-bh_0.$ On the other hand, the third expression becomes
$$h_2=h_1\left(\frac{g_1'}{g_0'}+\frac{f_1'}{f_0'}\right)-h_0\left(\frac{g_1'}{g_0'}\frac{f_1'}{f_0'}\right),$$
which implies
$$\frac{g_1'}{g_0'}+\frac{f_1'}{f_0'}=a,\quad \frac{g_1'}{g_0'}\frac{f_1'}{f_0'}=b.$$
Hence
$$\left(\frac{g_1'}{g_0'}\right)^2-a\left(\frac{g_1'}{g_0'}\right)+b=0$$
with roots
$$\alpha=\frac{a}{2}+\frac{\sqrt{a^2-4b}}{2},\quad \beta=\frac{a}{2}-\frac{\sqrt{a^2-4b}}{2},\quad a=\beta+\alpha, \quad  b=\alpha\beta.$$
Then
\begin{equation}\label{g_1f_1}
g_1=\alpha g_0+\gamma,\quad f_1=\beta f_0+\delta,
\end{equation}
where $\gamma$ and $\delta$ are constants.

\p Before going back to  the constraint in \eqref{h_0h_1I}, notice that the tentative solution \eqref{solutionsI_I} can be rewritten in a more compact formulation using only the functions
$g_0,$ $g_1,$ $f_0,$ $f_1,$ $h_0$ and $h_1,$ as
\begin{eqnarray}\label{solutionsIg0f0g1f1}
g_{k+1}&=&A_{k}(ag_1-bg_0+c)-A_{k-1}bg_1,\quad f_{k+1}=A_{k}(af_1-bf_0+d)-A_{k-1}bf_1,\nonumber\\
h_{k+1}&=&A_{k}(ah_1-bh_0)-A_{k-1}bh_1,\nonumber\\
A_{k+1}&=&aA_k-bA_{k-1},\quad A_{0}=0,\quad A_1=1,\qquad\qquad k=1,2,\dots.
\end{eqnarray}
Then, using \eqref{g_1f_1}, and the constant $\alpha,$ $\beta$ instead of $a,$ $b,$ these expressions become
\begin{eqnarray}\label{solutionsIg0f0}
g_{k+1}&=&\alpha(A_{k+1} - A_k \beta)g_0+A_{k+1}\gamma ,\quad f_{k+1}=\beta(A_{k+1} - A_k \alpha)f_0+A_{k+1}\delta,\nonumber\\
h_{k+1}&=&A_{k+1}h_1-\alpha\beta A_{k} h_0,\nonumber\\
A_{k+1}&=&(\alpha+\beta)A_k-\alpha\beta A_{k-1},\quad A_{0}=0,\quad A_1=1,\qquad \qquad k=1,2,\dots,
\end{eqnarray}
and the $h_0$ and $h_1$ functions in \eqref{h_0h_1I} are:
\begin{equation*}
h_0=f_0'g_0\alpha+f_0g_0'\beta+f_0'\gamma+g_0'\delta,\quad
2h_1=(\alpha+\beta)h_0+g_0'\alpha\delta+f_0'\beta\gamma,
\end{equation*}
where the constants of integrations, $c$ and $d,$ have been absorbed into the constants $\gamma$ and $\delta.$

\p Now it is time to analyse the constraint in \eqref{h_0h_1I}. It reads:
\begin{equation}\label{constraintI}
(\alpha- \beta)(\beta g_0'f_0+\alpha g_0f_0')+ f_0'\gamma\alpha-g_0'\delta\beta=0.
\end{equation}
There are two possibilities that will be explored in the next two subsections.

\subsubsection{Solution B$_1$}
\label{GeneralCaseIa}

\p If $\alpha=\beta$ the constraint \eqref{constraintI} simplifies to
\begin{equation}\label{zetaconstantg0f0}
f_0'\gamma=g_0'\delta,\quad \Rightarrow \quad f_0=\frac{\delta}{\gamma}\,g_0+\varepsilon,
\end{equation}
where $\varepsilon$ is a constant and the functions $h_0,$ $h_1$ become
\begin{equation}\label{zetaconstanth0h1}
h_0=g_0\,g_0'\,2\alpha \frac{\delta}{\gamma}+g_0'(2\delta+\alpha \varepsilon),\quad h_1=\alpha (h_0+g_0'\delta).
\end{equation}
Finally, it can be easily notice that $A_k$ in \eqref{solutionsIg0f0} can be rewritten as $A_k=k\,\alpha^{k-1}$ for all $k.$ Hence, expressions \eqref{solutionsIg0f0} simplify and
the solution B$_1$ is:
\begin{eqnarray}\label{solutionIa}
g_{k}&=&\alpha^k\,g_0+k\,\gamma\,\alpha^{k-1},\quad
f_{k}=\alpha^k\left(\frac{\delta}{\gamma}\,g_0+\varepsilon\right)+k\,\delta\alpha^{k-1},\nonumber \\
h_{k}&=&\alpha^k (h_0+k \,g_0'\delta),\qquad \qquad k=0,1,2,\dots
\end{eqnarray}
where $g_0$ is a free function of $q$ and $\alpha,$ $\delta,$ $\gamma,$ $\varepsilon$ are constants.
The claim is that this solution satisfies all the relations \eqref{generalrelationsfg}. Some details will be provided in appendix \ref{all_relations}.

\p Hence
\begin{equation*}
F(p+\lambda,q)=e^{(p+\lambda)\alpha/2}\left(\frac{\delta}{\gamma}g_0+\varepsilon+\delta\left(\frac{p+\lambda}{2}\right)\right),\ \  G(p-\lambda,q)=e^{(p-\lambda)\alpha/2}\left(g_0+\gamma\left(\frac{p-\lambda}{2}\right)\right),
\end{equation*}
and
$$F_\lambda G_q-F_q G_\lambda=\frac{e^{p\,\alpha}}{2}\left(g_0'g_0 \,2\alpha\frac{\delta}{\gamma}+g_0'(\alpha\varepsilon+2\delta+\alpha \,\delta \,p)\right)=U(p+q)-V(p-q).$$
If $\delta\ne 0$, the presence of the last term proportional to $p$ makes it hopeless to find suitable potentials. However, setting $\delta=0$ the previous expression becomes
$$F_\lambda G_q-F_q G_\lambda=e^{p\,\alpha}\,g_0'\frac{\alpha\, \varepsilon}{2}.$$
This suggests
$$g_0'\sim (e^{\alpha\,q}-e^{-\alpha\,q})\quad \Rightarrow \quad  U(u)=V(u)=\frac{\alpha\, \varepsilon}{2}\,e^{\alpha u},$$
or
$$g_0'\sim e^{\alpha\,q}\quad \Rightarrow \quad  U(u)=\frac{\alpha\, \varepsilon}{2}\,e^{\alpha u},\quad V=0.$$
In the first case, a field satisfying the Liouville potential is located on both sides
of the defect and in the second case there is a Liouville field on one side of the defect and a free massless field on the other.

\subsubsection{Solution B$_2$}
\label{GeneralCaseIb}

\p This expression \eqref{constraintI} can be rewritten as
$$\frac{g_0'}{f_0'}=\frac{\alpha g_0(\alpha- \beta)+\alpha \gamma}{\beta\delta-\beta f_0(\alpha- \beta)}\equiv \zeta,$$
where $\zeta$ is a function of $q.$ This leads to two equations that need to be solved
\begin{equation}\label{systemOfDofEqI}
g_0'=\zeta f_0',\quad g_0=\frac{\zeta \delta\beta-\alpha\gamma}{\alpha(\alpha- \beta)}-\frac{\beta}{\alpha}\,\zeta f_0.
\end{equation}
By differentiating the second equation it is found
$$\frac{f_0'(\alpha^2- \beta^2)}{\beta\delta-\beta f_0(\alpha-\beta)}=\frac{\zeta'}{\zeta},$$
which leads to
\begin{equation}\label{g_0f_0Ib}
f_0=-\frac{\varepsilon\zeta^{-\beta/(\alpha+\beta)}}{\beta(\alpha- \beta)}+\frac{\delta}{(\alpha- \beta)},\quad
g_0=\frac{\varepsilon\zeta^{\alpha/(\alpha+\beta)}}{\alpha(\alpha- \beta)}-\frac{\gamma}{(\alpha- \beta)},
\end{equation}
and
\begin{equation}\label{h_0h_1Ib}
h_0=-f_0'\frac{\beta \gamma}{(\alpha- \beta)}+g_0'\frac{\alpha \delta}{(\alpha- \beta)}\quad
h_1=-f_0'\frac{\beta^2 \gamma}{(\alpha- \beta)}+g_0'\frac{\alpha^2 \delta}{(\alpha- \beta)}.
\end{equation}
Looking at expressions \eqref{solutionsIg0f0}, it can be noticed that
$$A_{k+1}-\alpha A_k=\beta^{k+1},\quad A_{k+1}-\beta A_k=\alpha^{k+1},\quad A^k=\left(\frac{\alpha^k- \beta^k}{\alpha- \beta}\right),\qquad k=0,1,\dots$$
Clearly $\alpha\neq \beta,$ which is fine since the case $\alpha= \beta$ has been explored in the previous subsection.
Then the solution B$_2$ is:
\begin{eqnarray}\label{SolutionIb}
f_{k}&=&\beta^{k}f_0+\delta\,\left(\frac{\alpha^k- \beta^k}{\alpha- \beta}\right),\quad
g_{k}=\alpha^kg_0+\gamma\,\left(\frac{\alpha^k- \beta^k}{\alpha- \beta}\right),\nonumber\\
h_k&=&-f_0'\frac{\beta^{k+1} \gamma}{(\alpha- \beta)}+g_0'\frac{\alpha^{k+1} \delta}{(\alpha- \beta)},\qquad \qquad k=0,1,\dots
\end{eqnarray}
with $g_0$ and $f_0$ given in \eqref{g_0f_0Ib}.  Once again, all relations \eqref{generalrelationsfg} are satisfied. Details can be found in appendix \ref{all_relations}.
Hence
\begin{eqnarray*}
F(p+\lambda,q)&=&\frac{\delta}{(\alpha- \beta)}\,e^{\alpha(p+\lambda)/2}+\left(f_0-\frac{\delta}{(\alpha- \beta)}\right)e^{\beta(p+\lambda)/2}\\
G(p-\lambda,q)&=&\left(g_0+\frac{\gamma}{(\alpha- \beta)}\right)\,e^{\alpha(p-\lambda)/2}-\frac{\gamma}{(\alpha- \beta)}\,e^{\beta(p-\lambda)/2},
\end{eqnarray*}
\begin{eqnarray}\label{PB_relationIb}
F_\lambda G_q-F_q G_\lambda&=&g'_0\,\frac{\delta\,\alpha}{2(\alpha- \beta)}\,e^{\alpha p}-f'_0\,\frac{\gamma\,\beta}{2(\alpha- \beta)}\,e^{\beta p}\nonumber\\
&=&\frac{\varepsilon\,\zeta'\,z^{-\beta/(\alpha+\beta)}}{2(\alpha-\beta)^2(\alpha+\beta)}\left(\delta\alpha\,e^{\alpha p}-\gamma\beta\,\zeta^{-1}\, e^{\beta p}\right)=U(p+q)-V(p-q).\nonumber\\
\end{eqnarray}
Before looking at the most general solution, it is easy to see that a solution is provided by
$$\zeta\sim e^{\alpha+\beta}\quad \Rightarrow \quad  U(u)=\frac{\varepsilon\,\delta\,\alpha}{2(\alpha+\beta)^2}\,e^{\alpha u}\quad V(v)=\frac{\varepsilon\,\gamma\,\beta}{2(\alpha+\beta)^2}\,e^{\beta v}.$$
Notice that this suggests the possibility to have two Liouville potentials on the two sides of the defect with different and arbitrary normalisations.
This is not surprising given that no mass is involved and that a type II defect can be seen as the result of a two fused type I defects \cite{cz2010}.

\p The most general solution to \eqref{PB_relationIb} is instead obtained by setting
\begin{eqnarray}\label{zeta_forIb}
\frac{\varepsilon\,\alpha\,\delta}{2(\alpha-\beta)^2(\alpha+\beta)}\,\zeta'\,\zeta^{-\beta/(\alpha+\beta)}&=&A\,e^{\alpha q}-B\,e^{-\alpha q},\nonumber\\
\frac{-\varepsilon\,\beta\,\gamma}{2(\alpha-\beta)^2(\alpha+\beta)}\,\zeta'\,\zeta^{-1-\beta/(\alpha+\beta)}&=&C\,e^{\beta q}-D\,e^{-\beta q},
\end{eqnarray}
whose ratio suggests
$$-\frac{\alpha\,\delta}{\beta\,\gamma}\zeta=\left(\frac{A\,e^{\alpha q}-B\,e^{-\alpha q}}{C\,e^{\beta q}-D\,e^{-\beta q}}\right)\equiv \frac{X(q)}{Y(q)}.$$
Then, the first expression in \eqref{zeta_forIb} can be rewritten as follows
\begin{equation}\label{XY_relation}
c(X'Y-Y'X)X^{n-1}=Y^{n+2}\quad \mbox{with} \quad c\equiv\frac{\varepsilon\,\gamma^{n+1}}{2\,\delta^n\,\beta^2}\frac{n^{3+n}}{(2n+1)^2(n+1)^n},
\end{equation}
where $n=-\beta/(\alpha+\beta).$ This expression has solution only for $n=1$ and $n=-2$ (the cases $n=0$ and $n=-1$ are excluded since they imply $b=0$ and therefore  $H_B=0$). These two solutions lead to the same $U$ and $V$ potentials.
Note in fact that by sending $n$ to $-n-1,$ \eqref{XY_relation} becomes
$$c'(Y'X-X'Y)Y^{n-1}=X^{n+2},$$
where $c'$ is a constant, unimportant for the current discussion. Clearly, if \eqref{XY_relation} has a solution, then this expression has a solution as well and the two solutions are related by the swapping of $\alpha$ and $\beta.$
In the end, setting $n=1$ the most general solution to \eqref{zeta_forIb} is:
$$\zeta=\frac{18 \beta}{\varepsilon\,\gamma}\left(C\,e^{\beta q}+D\,e^{-\beta q}\right)$$
$$U(u)=C\,e^{\beta u}-\frac{36\,D^2\,\delta\,\beta}{\varepsilon \gamma^2}\,e^{-2\beta u}\quad V(v)=D\,e^{\beta v}-\frac{36\,C^2\,\delta\,\beta}{\varepsilon \gamma^2}\,e^{-2\beta v},$$
which correspond to the Tzitz\'eica potentials. Once again there is a freedom in the choice of the exponential coefficients, which reflects the freedom of shifting the fields $u$ and $v$ by arbitrary constants,
as it was the case for the sinh-Gordon potential in section \ref{h2k=0}.
Setting $\beta=-1,$ $C=D=2$ and $36\,\delta/\varepsilon\,\gamma^2=1/2$
$$U(u)=V(u)=2\,e^{-u}+e^{2u},$$
so that a more familiar form for the Tzitz\'eica potential is recovered.

\p Finally, note that the sinh-Gordon potential is not a solution of \eqref{PB_relationIb}. In fact, it corresponds to the case $\alpha=-\beta,$ which is explicitly excluded since it would imply $a=0$, and therefore $H_C=0.$

\section{Conclusion}

\p This article adds another piece in the complex mosaic that represents integrable field theory. That these models are special is a well known fact.
It turns out that they are the only relativistic field theories able to support a purely transmitting defect,
which is defined by the requirement of both energy and momentum conservation. All of this is achieved by preserving their most distinctive feature: integrability. Somehow, demanding both energy and momentum conservation
singles out the integrable models. Previously known results concerning the sinh-Gordon,
Tzitz\'eica and Liouville models have been recovered.
In addition, it was interesting to see how the possibility to have two differently normalised Liouville models on either side of the defect emerges naturally from the current investigation.
This is a possibility that, though not surprising, was not explicitly considered previously.

\p Nevertheless, the present investigation has a limitation. Only models with a single scalar field have been considered. Previously, multi-scalar field theories supporting type I defects have been analysed and found to be the non-affine and affine Toda fields models based on the $a_n^{(1)}$ root data \cite{bcz2004,cz2007}. In order to extend the more general investigation to multiple scalar field models, it is useful to borrow an
idea presented recently in \cite{br2017}, then applied successfully to the $d_r^{(1)}$ affine Toda models, to mix the type I and type II defects in order
to increase the range of models that support these kinds of defect and hopefully to demonstrate that this will allow all integrable Toda models. The most general sewing conditions for the type I defects can be found in
\cite{bcz2004,cz2007} and are:
$$u_x=Au_t+(1-A)v_t-{\cal E}_u,\quad v_x=-Au_t+(1+A)u_t+{\cal E}_v,$$
where the fields $u$ and $v$ are now vectors representing multi component scalar fields and $A$ is an antisymmetric matrix.
On the other hand the sewing conditions for the type II defect look unchanged with respect to the ones seen previously. In fact they are:
$$u_x=\lambda_t- {\cal E}_u,\ \ v_x=\lambda_t+{\cal E}_v,\ \ u_t-v_t =-\ce_\lambda,$$
where $u$ $v$ and $\lambda$ are now vectors.
The mixing idea consists in splitting the space in which the fields live into two pieces. The fields belonging to one part will satisfy type I sewing conditions at the defect and the fields belonging to the other part will
satisfy type II sewing conditions. In order to keep track of this aspect, it is convenient to introduce two projection operators, $\Gamma_1$ and $\Gamma_2$ such that $\Gamma_1+\Gamma_2=1, \ \Gamma_k^2=\Gamma_k,\ k=1,2.$
Momentum conservation leads to the following constraints:
\begin{eqnarray*}
&&{\cal E}_{(\Gamma_1 \,q)}+2A{\cal E}_{(\Gamma_1\, p)}=-{\cal P}_{(\Gamma_1\, q)},\quad {\cal E}_{(\Gamma_1\, p)}={\cal P}_{(\Gamma_1\, p)},\\
&&{\cal E}_{(\Gamma_2 \,p)}={\cal P}_{(\Gamma_2 \,\lambda)},\quad {\cal E}_{(\Gamma_2 \,\lambda)}={\cal P}_{(\Gamma_2 \,p)},\\
&& \frac{1}{2}\left({\cal E}_{(\Gamma_2\, q)}{\cal P}_{(\Gamma_2 \,\lambda)}-{\cal E}_{(\Gamma_2\, \lambda)}{\cal P}_{(\Gamma_2\, q)}+{\cal E}_{(\Gamma_1\, p)}{\cal E}_{(\Gamma_1\, q)}\right)=U-V,
\end{eqnarray*}
where the usual definition $p=(u+v)/2,$ $q=(u-v)/2$ have been used. Note that the subscripts in parentheses indicate derivatives.
It is useful to introduce a new field variable $\xi=-A\Gamma_1 \,q.$ Then the previous constraints become
\begin{eqnarray*}
&&{\cal E}_{(\Gamma_1 \,\xi)}-2{\cal E}_{(\Gamma_1\, p)}=-{\cal P}_{(\Gamma_1\, \xi)},\quad {\cal E}_{(\Gamma_1\, p)}={\cal P}_{(\Gamma_1\, p)},\\
&&{\cal E}_{(\Gamma_2 \,p)}={\cal P}_{(\Gamma_2 \,\lambda}),\quad {\cal E}_{(\Gamma_2 \,\lambda)}={\cal P}_{(\Gamma_2 \,p)},\quad {\cal E}_{(\Gamma_2\, \xi)}=-{\cal P}_{(\Gamma_2\, \xi)},\\
\end{eqnarray*}
which imply
\begin{eqnarray*}
{\cal E}&=&F(\Gamma_2 (p+\lambda), \Gamma_2 q, \Gamma_1(p+\xi))+G(\Gamma_2 (p-\lambda), \Gamma_2 q, \Gamma_2\xi),\\
{\cal P}&=&F(\Gamma_2 (p+\lambda), \Gamma_2 q, \Gamma_1(p+\xi))-G(\Gamma_2 (p-\lambda), \Gamma_2 q, \Gamma_2\xi)
\end{eqnarray*}
and
\begin{equation}\label{newPBrelation}
{F}_{(\Gamma_2\, \lambda)}{G}_{(\Gamma_2 \,q)}-{G}_{(\Gamma_2\, \lambda)}{F}_{(\Gamma_2 \,q)}-\frac{1}{2}{F}_{(\Gamma_1\, p)}\Gamma_1 A{F}_{(\Gamma_1\, \xi)}-\frac{1}{2}{F}_{(\Gamma_1\, p)}\Gamma_2 A{F}_{(\Gamma_2\, \xi)}=U-V.
\end{equation}
It can be seen how the first two terms on the left hand side are similar to the terms in the Poisson-Bracket relation \eqref{PBrel} investigated in the present article. On other hand, the other two terms take into account a mixing represented by the fields
$\xi,$ which are not confined to a single subspace. In \cite{br2017}, solutions for an expression similar to \eqref{newPBrelation} have been found. However, a complete set of solutions is still missing.

\appendix
\section{On the general expressions for solutions B}
\label{all_relations}

\p It was stated that solutions presented in section \eqref{GeneralCaseI} satisfy all relations \eqref{generalrelationsfg}. It suffices to look at one set of these equations. Consider, for instance, the first set in the first line.
For $k=0$ using \eqref{solutionsIg0f0g1f1} it is found that
$$f_1=\frac{h_1g_1-h_0(ag_a-bg_0+c)}{g_1'g_1-g_0'(ag_a-bg_0+c)}.$$
Notice that the $s$ and $r$ dependence disappears. This relation is easily seen to be satisfied using \eqref{h_0h_1I}.
For $k>0$ it is convenient to rewrite the relations using \eqref{solutionsIg0f0} then
\begin{eqnarray}\label{generalfkplus1}
f_{k+1}&=&\frac{h_1g_0 X_{k+1}+h_1 Y_{k+1}-h_0g_0 bX_k-h_0 bY_k}{-g_0'\beta\gamma}, \nonumber \\
X_{k+2}&=&(\alpha+\beta)X_{k+1}-\alpha\beta X_{k},\quad X_1=\alpha,\quad X_2=\alpha\beta,\nonumber \\
Y_{k+2}&=&(\alpha+\beta)Y_{k+1}-\alpha\beta Y_{k},\quad Y_1=\gamma,\quad Y_2=0,\qquad\qquad k=1,2,\dots
\end{eqnarray}
Before proceeding any further, note that because these relations are obtained using \eqref{solutionsIg0f0g1f1} or \eqref{solutionsIg0f0}, the $r$ and $s$ must be greater than $1$.
However, the cases in which $r$ and/or $s$ are $0$ or $1$ have already been considered since they were used in section \ref{generalcase} in order to get expressions
\eqref{GeneralRelations_hg_firstline}, \eqref{GeneralRelations_hg_secondline}.

\p Consider first solution B$_1$. Because $\alpha=\beta,$ the constants appearing in \eqref{generalfkplus1} simplify. They become
$$X_{k}=\alpha^k,\quad Y_{k}=-\gamma(k-2)\alpha^{k-1},\quad k=1,2,\dots.$$
Using these expressions in \eqref{generalfkplus1} together with expressions \eqref{zetaconstanth0h1} for $h_0$ and $h_1,$ it is not difficult to see that
\eqref{generalfkplus1} reproduces the expressions for $f_{k+1}$ in \eqref{solutionIa}.

\p Consider now the solution B$_2$. Using \eqref{h_0h_1Ib} expressions \eqref{generalfkplus1} become
\begin{eqnarray}\label{generalfkplus1b}
f_{k+1}&=&\frac{-\gamma\, f_0'\,\beta^2}{(\alpha-\beta)(-g_0'\beta\gamma)}\left(g_0(X_{k+1}-\alpha X_k)+(Y_{k+1}-\alpha Y_k)\right)\nonumber\\
&&+\frac{\delta\, g_0'\,\alpha^2}{(\alpha-\beta)(-g_0'\beta\gamma)}\left(g_0(X_{k+1}-\beta X_k)+(Y_{k+1}-\beta Y_k)\right).
\end{eqnarray}
It is not difficult to realise that
\begin{eqnarray*}
X_{k+1}-\alpha X_k&=&-\alpha\,(\alpha-\beta)\,\beta^{k-1},\quad X_{k+1}-\beta X_k=0,\\
Y_{k+1}-\alpha Y_k&=&-\alpha\gamma\,\beta^{k-1},\quad  Y_{k+1}-\beta Y_k=-\beta\gamma\,\alpha^{k-1},\quad k=1,2\dots
\end{eqnarray*}
Using these relations into \eqref{generalfkplus1b}, expressions \eqref{SolutionIb} for  $f_{k+1}$ are recovered.

\section{The $H_A=0$ case}
\label{H_A=0}

\p As an example, assume $H_A,$ in \eqref{FGfunctions} is zero.
This implies, for instance, that $G_A=0.$
Then, since the denominator of the expressions in the first line of \eqref{generalrelationsfg} for  $(r,s)=(0,1)$ is $G_A,$ it must be
$$h_{k}\,g_2-h_{k+1}\,g_1=0\quad h_{k+1}\,g_0'-h_k\,g_1'=0.$$
Provided the $g$-functions are different from zero and $g_0'$ and $g_1'$ are also different from zero,
\footnote{The case in which $g_0'$ and $g_1'$ are zero produces a $G$ function that is $q$ independent. This does not lead to new solutions.} it follows that
\begin{equation}\label{sc_h}
\frac{h_{k+1}}{h_k}=\frac{g_2}{g_1}=\frac{g_1'}{g_0'},\quad\Rightarrow \quad  h_{k+1}=h_k\,\frac{g_1'}{g_0'}=h_0\left(\frac{g_1'}{g_0'}\right)^{k+1},\quad k=0,1,\dots.
\end{equation}
On the other hand  using formulas in the second line of \eqref{generalrelationsfg}, it is found that
$$\frac{g_{k+1}}{g_1}=\frac{h_{k+r}f_{s+1}-h_{k+s}f_{r+1}}{h_{r}f_{s+1}-h_{s}f_{r+1}}=\left(\frac{g_1'}{g_0'}\right)^k,\quad\mbox{and}\quad
\frac{g_{k'}}{g_0'}=\left(\frac{g_1'}{g_0'}\right)^k,\quad k=0,1,\dots$$
where \eqref{sc_h} has been used.
The compatibility condition reads
$$g_0'\left(\frac{g_1'}{g_0'}\right)^{k}=g_1'\left(\frac{g_1'}{g_0'}\right)^{k-1}+g_1\,(k-1)\left(\frac{g_1'}{g_0'}\right)'\left(\frac{g_1'}{g_0'}\right)^{k-2}\quad k=1,2,\dots,$$
that is
$$\frac{g_1}{g_0'}\left(\frac{g_1'}{g_0'}\right)'=0,\quad \Rightarrow\quad \left(\frac{g_1'}{g_0'}\right)'=0,$$
sice $g_1$ is different from zero.
Then $g_1'=c\,g_0',$
where $c$ is a constant, and $g_1=cg_0+\alpha.$ It follows
\begin{equation}\label{HzeroRelationsFor_g_h}
g_{k+1}=g_0\,c^{k+1}+\alpha c^k,\quad h_{k+1}=h_0\,c^{k+1}\quad k=0,1,\dots.
\end{equation}
It is easy to verify that $H_B$ and $H_C$ are also zero.

\p Looking at the general formula \eqref{fghrelsRearranged}, it is possible to notice that the ratio $g'_{N-k}/g_{N-k+1}$ on the right hand side are all the same and equal to $g_0'/g_1$ that is:
$$\frac{h_N}{f_{k+1}g_{N-k+1}}= \left(\frac{g'_0}{g_1}\right)+\left(\frac{f'_{k}}{f_{k+1}}\right)\quad k=0,1,\dots.$$
Using \eqref{HzeroRelationsFor_g_h} this leads to
$$h_0=\frac{f_{k+1}\,g_0'+f_k'\,g_1}{c^k}=f_1g_0'+f_0'g_1, $$
that is
\begin{equation}\label{HA0_frelations}
(f_{k+1}-c^kf_1)\,g_0'+(f_k'-c^kf_0')\,g_1=0.
\end{equation}
Note that the same expression can also be found by using the expressions in the second line of \eqref{generalrelationsfg} with, for instance, $(r,s)=(0,s).$
A solution is
$$f_{k+1}=c^k\,f_1,\quad f_k'=c^k\,f_0' \quad \Rightarrow \quad f_1=c\,f_0+\beta.$$
Note that different solutions to \eqref{HA0_frelations} do not translate into different expressions for the final $U$ and $V$ potentials. In fact, their forms depend on $(F_\lambda G_q-F_q G_\lambda),$ which,
because of \eqref{HzeroRelationsFor_g_h} is proportional to $h_0\,e^{cp}.$ A different choice of $f-$functions would have an effect on $h_0,$ which, in any case,
must be proportional to $A\,e^{cq}-B\,e^{-cq}$ where $A$ and $B$ are arbitrary constants.
In summary:
$$g_{k+1}=g_0\,c^{k+1}+\alpha c^k,\quad f_{k+1}=f_0\,c^{k+1}+\beta c^k,\quad h_{k+1}=(c(g_0f_0)+\alpha f_0+\beta g_0)'\,c^{k+1}\quad k=0,1,\dots.$$
Then
$$F(p+\lambda,q)=e^{(p+\lambda)c/2}f_0+\frac{\beta}{c}\left(e^{(p+\lambda)c/2}-1\right),\quad G(p-\lambda,q)=e^{(p-\lambda)c/2}g_0+\frac{\alpha}{c}\left(e^{(p-\lambda)c/2}-1\right),$$
and
$$F_\lambda G_q-F_q G_\lambda=\frac{e^{cp}}{2}(c(g_0f_0)+\alpha f_0+\beta g_0)'.$$
This implies
$$(c(g_0f_0)+\alpha f_0+\beta g_0)'\sim \left(A\,e^{cq}-B\,e^{-cq}\right)\quad \Rightarrow \quad  U(u)=\frac{A}{2}\,e^{c u},\quad U(u)=\frac{B}{2}\,e^{c v}.$$
According to the values of the constants $A$ and $B$ there is a Liouville field on both sides of the defect or a Liouville field on one side and a free massless field on the other.
A similar analysis can be performed for the cases $H_B=0$ and $H_C=0.$

\end{document}